\documentclass[pra,aps,twocolumn,byrevtex,showpacs,floatfix,tightenlines,lengthcheck]{revtex4}

\usepackage{epsfig}

\begin{document}
\DeclareGraphicsExtensions{.eps, .jpg}

\title{Creation of maximally entangled photon-number states using optical fiber multiports}
\author{G. J. Pryde}
\email{pryde@physics.uq.edu.au}
\author{A. G. White}
\email{andrew@physics.uq.edu.au}
\affiliation{Centre for Quantum Computer Technology and Department of Physics, University of Queensland, Brisbane, QLD 4072, Australia.}

\date{1 May, 2003}

\begin{abstract}
We theoretically demonstrate a method for producing the maximally path-entangled state $(1/\sqrt{2})\hspace{2pt}\left(|N,0\rangle+\exp[iN\phi]|0,N\rangle\right)$ using intensity-symmetric multiport beamsplitters, single photon inputs, and either photon-counting postselection or conditional measurement. The use of postselection enables successful implementation with non-unit efficiency detectors. We also demonstrate how to make the same state more conveniently by replacing one of the single photon inputs by a coherent state.
\end{abstract}
\pacs{03.67.Mn, 42.50.Dv, 42.50.St,  06.30.Bp}
\maketitle

%\vspace{-1 cm}
There is great interest in the development of a range of nonclassical optical states for quantum technology applications \cite{dowlingmilburn}, including applications in quantum computing \cite{klm}; quantum cryptography \cite{gisin}; quantum metrology and lithography \cite{dowlingrosetta,itoh,fonseca,bollinger,ralph,boto,shihlithog}; and quantum imaging \cite{imag}. A  development of particular interest is the ``quantum Rosetta stone'' \cite{dowlingrosetta}, which elucidates the isomorphism between Mach-Zehnder interferometers (MZIs), Ramsey spectroscopy and quantum phase gates. It has been shown theoretically that a MZI produces interferometric fringes with a period $\lambda/N$ when employed with a maximally path-entangled $N$-photon state,
\begin{equation}
\label{eq:noonstate}
{|\psi\rangle}_{\text{NOON}}=\frac{1}{\sqrt{2}}(|N,0\rangle + e^{i N\phi}|0,N\rangle),
\end{equation}
within the interferometer, where the ordered pair in the kets represents the photon number in each of the two interferometer arms. The period is therefore 1/$N$ times smaller than that for a MZI employed with a coherent state or single photon input of wavelength $\lambda$ \cite{bollinger}. This has been experimentally verified for $N=2$ \cite{itoh}, and an equivalent result obtained in Ramsey spectroscopy \cite{meyer}. These reduced-period fringes can be used to measure a phase shift in one arm of the interferometer with a phase uncertainty $\Delta\phi = 1/N$ at the Heisenberg limit for all $N$ and $\phi$ \cite{bollinger}. The same quantum state ${|\psi\rangle}_{\text{NOON}}$ has been suggested for use in generating spatial interferometric fringes with a period $N$ times smaller than single photon fringes, with potential application to beating the diffraction limit in lithography \cite{boto}. The reduced period of a spatial interference pattern has been demonstrated for the case $N=2$ \cite{shihlithog}. There is significant interest in the extension of these techniques to larger $N$, and there exist a number of proposals for achieving this, each of which is experimentally intensive \cite{fiurasek,kok,zou,gerry}. 

\begin{figure}[!htb]
\center{\epsfig{figure=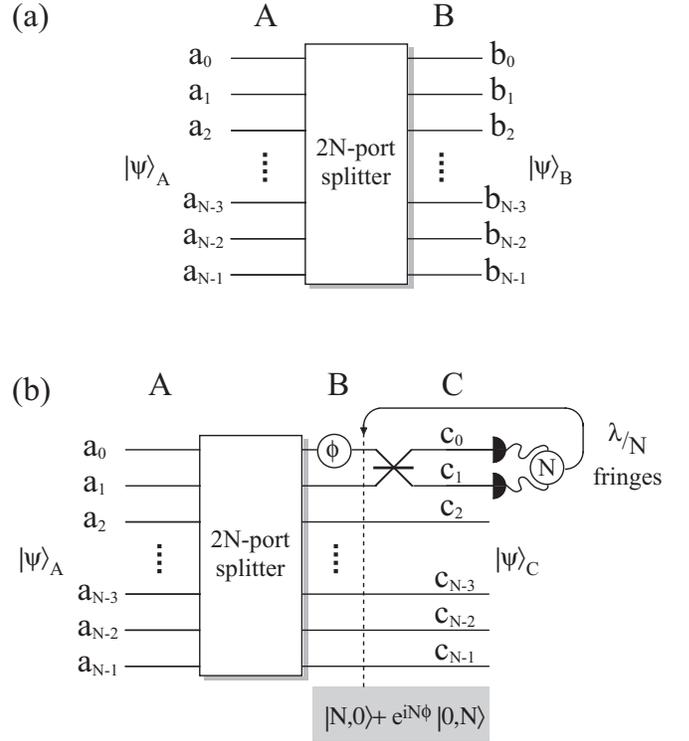,width=\columnwidth}}
\caption{(a) Conceptual diagram of $2N$-port intensity-symmetric beamsplitter, with input and output modes labeled. (b) Mach-Zehnder interferometer, for measuring phase shift $\phi$, created with a $2N$-port splitter and a 50/50 beamsplitter. Postselection occurs when the two detectors count a total of $N$ photons between output modes 0 and 1. The postselection guarantees that the state at B was the maximally number entangled state, and fringes with a period 1/$N$ times the single photon wavelength are observed in N-photon coincidence.}
\label{fig:multiport}
\end{figure}

In this paper, we theoretically demonstrate a method of producing the quantum state ${|\psi\rangle}_{\text{NOON}}$, for general $N$, by means of a $2N$-port ($N$ inputs, $N$ outputs) beamsplitter \cite{note2N,mattle,zukowski} with symmetric splitting ratios, and photon counting. The technique we present is less intensive, in that it uses a single beamsplitting device instead of  a cascaded array of beamsplitter and measurement stages, and there is correspondingly a significant reduction in the number of optical elements required. Further, it has the advantage of operating successfully with non-unit efficiency detectors because it employs postselection.  An intensity-symmetric $2N$-port beamsplitter is readily implemented in practice with a $N\times N$ fiber coupler, which are available commercially for various values of $N$ up to $N=32$. A further development is the simplification of input state preparation by replacing one of the single photon inputs with a coherent state. 

The $2N$-port itself is represented in Fig.~\ref{fig:multiport}(a), where it is employed with $N$ single photon inputs. The input state can be written:
\begin{equation}
\label{eq:nitterinput}
{|\psi\rangle}_A=\left[\prod_{k=0}^{N-1}\hat{a}^\dag_k\right]\hspace{4pt}{|0_0,\ldots,0_{N-1}\rangle}_A.
\end{equation}
To obtain the output state, we act on the input state with the $2N$-port transformation operator, and use its unitarity and the time reversal symmetry of the network to write:
\begin{eqnarray}
\label{eq:nitteroutstate}
\nonumber{|\psi\rangle}_B&=&\hat{U}_{\text{2N}}\hspace{2pt}{|\psi\rangle}_A\\
\nonumber&=&\hat{U}_{\text{2N}}\left[\prod_{k=0}^{N-1}\hat{a}^\dag_k\right]\hspace{4pt}{|0_0,\dots,0_{N-1}\rangle}_A\\
\nonumber&=&\hat{U}_{\text{2N}}\left[\prod_{k=0}^{N-1}\hat{a}^\dag_k\right]\hat{U}_{\text{2N}}^\dag\hat{U}_{\text{2N}}\hspace{4pt}{|0_0,\dots,0_{N-1}\rangle}_A\\
&=&\left[\prod_{k=0}^{N-1}\hat{a}^\dag_k(\hat{b}_0^\dag,\dots,\hat{b}_{N-1}^\dag)\right]\hspace{4pt}{|0_0,\dots,0_{N-1}\rangle}_B,
\end{eqnarray}
where $\hat{a}^\dag_k(\hat{b}_0^\dag,\dots,\hat{b}_{N-1}^\dag)$, abbreviated $\hat{a}^\dag_k(\hat{b}^\dag)$ henceforth, is the input operator for the $k$th mode written in terms of the output operators $\hat{b}_m^\dag$. In the case of the \emph{canonical} $2N$-port, the $\hat{b}^\dagger_k$ are related to the input bosonic mode operators $\hat{a}^\dagger_k$ by the matrix equation,
\begin{equation}
\label{eq:matrixeqn}
\widetilde{\hat{a}^\dag}(\hat{b}^\dag)=\mathbf{M}^\dag_{2N\text{,can}}\hspace{2pt}\widetilde{\hat{b}^\dag},
\end{equation}
where the canonical $2N$-port matrix elements are given by the discrete Fourier transform \cite{mattle},
\begin{equation}
\label{eq:matrixelements}
\mathbf{M}_{2N\text{,can}}^{(k+1,l+1)}=\frac{1}{\sqrt{N}}e^{2\pi i kl/N},
\end{equation}
and $k,l$ vary between $0$ and $N-1$. As an example, the canonical 6-port matrix is
\begin{equation}
\label{M3}
\mathbf{M_{\text{6,can}}}=\frac{1}{\sqrt{3}}\left( \begin{array}{ccc}
     1 & 1 & 1\\
      1 & e^{2\pi i/3}  & e^{4\pi i/3}\\
      1 & e^{4\pi i/3} & e^{2\pi i/3} 
\end{array} \right),
\end{equation}
and the output state, with single photon inputs, is
\begin{eqnarray}
\label{tritterstate}
\nonumber { |\psi_3\rangle}_{B} & =&\frac{1}{3\sqrt{3}}(\hat{b}_0^{\dag 3}+\hat{b}_1^{\dag 3}+\hat{b}_2^{\dag 3}-3\hat{b}_0^{\dag}\hat{b}_1^{\dag}\hat{b}_2^{\dag})\hspace{4pt}|0,0,0\rangle \\
\nonumber &=&\frac{\sqrt{2}}{3}(|3,0,0\rangle+|0,3,0\rangle+|0,0,3\rangle) \\ & &-\frac{1}{\sqrt{3}}|1,1,1\rangle.
\end{eqnarray}
This state is itself highly entangled. It is also of a form that is somewhat suggestive of the ${|\psi\rangle}_{\text{NOON}}$ state. In general, to obtain the ${|\psi\rangle}_{\text{NOON}}$ state from that of Eq.~(\ref{eq:nitteroutstate}), we invoke postselection using number-resolving photon counters \cite{takeuchi}, as shown in Fig.~\ref{fig:multiport}(b). The postselection occurs when the counters together count a total of $N$ photons simultaneously. Due to the requirement to have a total of $N$ photons in output modes 0 and 1, the only terms retained in the expansion of Eq.~(\ref{eq:nitteroutstate}) are those with operator products of the form $\hat{b}_0^{\dag N-k}\hat{b}_1^{\dag k}$, with $0\leq k\leq N$. It is important to note that the counting of $N$ photons must occur after the final beamsplitter, otherwise the states $|N,0\rangle$ and $|0,N\rangle$ are distinguishable, because the ``which-path'' information is accessible. In principle, it is possible to replace the MZI with other experiments that utilize ${|\psi\rangle}_{\text{NOON}}$, as long as the path information is not accessible. 

The postselection, performed as shown in Fig.~\ref{fig:multiport}(b), ensures that we know with certainty that whenever the detectors count a total of $N$ photons in modes $c_0$ and $c_1$, there are no photons in modes $c_2,\ldots,c_{N-1}$ and the state at $B$, the output of the $2N$-port itself, is known to have been  
\begin{eqnarray}
\label{eq:psstate}
\nonumber {|\psi\rangle}_{B}^{PS} & =&{|\psi_{01}\rangle}_{B}^{PS}\otimes |0_2,\ldots,0_{N-1}\rangle \\
\nonumber {|\psi_{01}\rangle}_{B}^{PS} & =&K\hspace{1pt}N^{-N/2}\left[\prod_{m=0}^{N-1}(\hat{b}^\dag_0+e^{2\pi im/N} \hat{b}^\dag_1)\right]\hspace{3pt}|0_0,0_1\rangle \\
\nonumber & =&K\hspace{1pt}N^{-N/2}(\hat{b}^{\dag N}_0 - (-1)^N \hat{b}^{\dag N}_1)\hspace{4pt}|0_0,0_1\rangle \\
 & =&K\hspace{1pt}\sqrt{(N!)}N^{-N/2}(|N,0\rangle-(-1)^N|0,N\rangle),
\end{eqnarray}
since the coefficients of the cross terms between $\hat{b}_0^\dag$ and $\hat{b}_1^\dag$ vanish, as proved in the Appendix. The postselection ensures that this technique works, even with non-unit efficiency number resolving counters, since at most $N$ counts are possible, and all counts less than $N$ are rejected. The state ${|\psi\rangle}_{B}^{PS}$ is normalized by the coefficient $K=N^{N/2}/(\sqrt{(N!)}\sqrt{2})$, and using Stirling's approximation in the case of large $N$ yields $1/K\approx\sqrt{\pi N}e^{-N}$. The latter gives the scaling law for the production of the maximally path-entangled $N$-photon state using our method. This exponential scaling law is similar to the scaling laws for other proposed methods \cite{fiurasek,zou,kok} of generating the state ${|\psi\rangle}_{\text{NOON}}$. 

Our technique can also be employed to prepare ${|\psi\rangle}_{\text{NOON}}$ \emph{conditionally} using projective measurement as opposed to postselection. The condition of $N$ photons in the output modes $b_0,b_1$ of the multiport is enforced by measuring vacuum on each of modes $b_2,\ldots,b_{N-1}$, instead of counting photons at the output of the MZI. Conditioning on the measurement of zero photons has the obvious disadvantage of requiring unit efficiency detectors, but with the advantage of removing the destructive measurement on modes 0 and 1.

The non-postselected output state of the complete MZI can be calculated using the time-reversed creation operators as before, 
\begin{equation}
\label{eq:mzinonps}
{|\psi\rangle}_{C}=\left[\prod_{k=0}^{N-1}\hat{a}^\dag_k(\hat{c}_0^\dag,\dots,\hat{c}_{N-1}^\dag)\right]\hspace{4pt}|0_0,\dots,0_{N-1}\rangle,
\end{equation}
but now the unitary operator is the transfer operator for the entire system, represented by the $N\times N$ matrix $\mathbf{M}^\dag_{\text{2N,can}}\mathbf{\Phi}^\dag\mathbf{M}^\dag_{\text{4,ex}}$, composed of the transfer matricies of the different elements of the optical network. The matrix $\mathbf{\Phi}$ represents the phase shift $\phi$ in Fig.~\ref{fig:multiport}(b) and is the identity except for the first diagonal element which is $e^{i\phi}$. The matrix $\mathbf{M}_{\text{4,ex}}$ represents the final beamsplitter in the MZI and is the identity except for the four elements in the top left corner which are the matrix elements for the canonical 4-port, as per Eq.~(\ref{eq:matrixelements}). Using the same proof as before, the (unnormalized) postselected state is 
\begin{equation}
\label{eq:fullmzips}
{|\psi\rangle}_{C}^{PS}=\left\{e^{-iN\phi}\left(\hat{c}_0^\dag+\hat{c}_1^\dag\right)^N-(-1)^N\left(\hat{c}_0^\dag-\hat{c}_1^\dag\right)^N\right\}\hspace{4pt}|0,0\rangle.
\end{equation}  
The form of this state agrees with the output state predicted for a Ramsey spectrometer or MZI employed with the state ${|\psi\rangle}_{\text{NOON}}$ \cite{bollinger}. Following Bollinger \emph{et al.} \cite{bollinger}, the phase measurement is made by measuring the parity of the count in the second detector, $(-1)^{N_{c_1}}$, which are the eigenvalues of the observable $\hat{O}$: 
\begin{equation}
\label{eq:obs}
\hat{O}(\hat{a}_0^\dag\pm\hat{a}_1^\dag)^N\hspace{2pt}|0,0\rangle=(\hat{a}_0^\dag\mp\hat{a}_1^\dag)^N\hspace{2pt}|0,0\rangle,
\end{equation} 
and which is equivalent to applying the Pauli $\sigma_z$ operator to each of the $N$ photons. The expectation value of $\hat{O}$ is then $\langle\hat{O}\rangle_{C}^{PS}=(-1)^N\cos(N\phi)$. Clearly the period of the interference is reduced by a factor of 1/$N$ over single photon interferometric measurement of a phase shift $\phi$, as expected for the maximally path-entangled state. It is straightforward to calculate the variance and deduce that the phase uncertainty is $1/N$ \cite{bollinger}, which is at the Heisenberg limit for the subensemble of postselected events.

We now consider the application of the multiport fiber beamsplitter in the case where one input state is replaced by a coherent state. Using a modified version of Eq.~(\ref{eq:nitteroutstate}) with one Fock state replaced by a displacement operator, the output state of the $2N$-port at point $B$, in the absence of postselection, is
\begin{eqnarray}
\label{eq:Ndimcoherent}
\nonumber|\psi_\alpha\rangle_{B}&=&\hat{D}_0(\alpha/\sqrt{N}) \\
\nonumber& &\hspace{4pt}\times\hspace{2pt}\left[\prod_{m=1}^{N-1}(\hat{D}_m(\alpha/\sqrt{N})\sum_{p=0}^{N-1}\hat{b}^\dag_p e^{2\pi imp/N})\right] \\
& &\hspace{8pt}\cdot\hspace{3pt}{|0_0,\dots,0_{N-1}\rangle}_B
\end{eqnarray}
where $\hat{D}_k(\alpha)$ is the coherent displacement operator acting on mode $k$. 
Postselection by counting $N$ photons in the first two output modes, as before, produces the state ${|\psi\rangle}_{\text{NOON}}$ in the limit of small $\alpha$, i.e. where $\alpha^2$ is much smaller than all the other photon input rates in the system. To see this, observe that the Fock inputs can at most contribute $N-1$ photons to the first two outputs. In general, the $N-1$ single photon inputs contribute $N-1-q$ (with $0\leq q \leq N-1$) photons to the first two output modes. and $q$ photons distributed somehow amongst the other output modes. The coherent state input will be required to contribute $q+1$ photons to the first two output modes to make a total of $N$ photons in those modes, and trigger the postselection. The largest amplitude term(s) of the state with $N$ photons in the first two modes goes as $\alpha^{q+1}$. In the small $\alpha$ limit, we discard terms with $q>0$ and the product of coherent state operators in Eq.~(\ref{eq:Ndimcoherent}) becomes
\begin{equation}
\label{eq:cohapprox}
\prod_{m=0}^{N-1}\hat{D}_m\approx\frac{\alpha}{\sqrt{N}}(\hat{b}^\dag_0+\hat{b}^\dag_1).
\end{equation}  
The other terms of order $\alpha$, i.e. $\alpha$ $\hat{b}^\dag_p$ ($1 < p\leq N-1$) are lost in the postselection because they do not contribute any photons to the first two output modes. With the approximation (\ref{eq:cohapprox}), Eq.~(\ref{eq:Ndimcoherent}) collapses to (\ref{eq:psstate}) and the maximally entangled state is obtained. 

There exists an \emph{exact} method (that does not constrain $\alpha$) for realizing this same transformation $|1,1,\dots,\alpha\rangle\rightarrow{|\psi\rangle}_{\text{NOON}}$, involving the use of photodetectors on all other outputs of the network of Fig.~\ref{fig:multiport}(b), as well as the first two. When each of the detectors on output ports 2 to $N-1$ registers \emph{no count}, simultaneously with the usual counting of $N$ photons in the first two output modes, the approximate relation (\ref{eq:cohapprox}) becomes exact. This is because all terms of order 2 or higher in $\alpha$ create at least one photon in an output mode other than 0 or 1, and so are postselected away. 

An interesting research problem is the replacement of \emph{more than one} of the Fock state inputs with coherent states. This is potentially a rich field of investigation, because of many potential phase and amplitude relationships that could be chosen between coherent input states. An open question is the effect of non-unit efficiency number resolving detectors when employing our technique with one or more coherent states, as possibility of having more than $N$ photons in the system potentially alters the postselection reliability. A starting point for such a study would be the detection model in Ref.~\cite{kokbraunstein}.
      
There are various important considerations for the practical application of the postselected $2N$-port scheme we have presented. The first is that it can be built for $N \leq 4$ with current technology. Down conversion sources capable of producing 4 single photons are becoming common and production rates are increasing \cite{jennewein}; number counting photon detectors are also becoming available \cite{takeuchi}. For small $N$, it is practical to replace the number-resolving detectors with beamsplitters and ordinary non-resolving photon counters, with slightly reduced overall probability of a postselection count. Taking $N=3$, for example, the number-counting detectors can be replaced with a non-resolving detector on output 0 and a 50/50 beamsplitter followed by two such detectors on output 1. The postselection occurs on a triple coincidence, and from Eq.~(\ref{eq:fullmzips}) it follows that the probability of such a detection still varies with $\phi$ as $\cos(N\phi)$.

We have already mentioned the commercial availability of 8-port and higher fibre splitters. This, however, leads to an important consideration. For $N \geq 4$, the general $2N$-port device is not necessarily described by the canonical multiport matrix, Eq.~(\ref{eq:matrixeqn}) \cite{mattle,bernstein}. For $N \leq 3$, the conservation of energy requirement defines the $2N$-port matrix to within external phases on the input and outputs, i.e. the canonical matrix entirely represents the physics for $N\leq 3$ \cite{mattle}. However, when $N\geq 4$, there exist free internal phases independent of the conservation of energy. These free phases lead to substantive changes in the 2$N$-port matrix. As an example, the matrix for $N=4$, with free internal phase $\theta$, can be written:
\begin{equation}
\label{m4theta}
\mathbf{M}_{\text{8,}\theta}=\frac{1}{2}\left( \begin{array}{cccc}
     1 & 1 & 1 & 1\\
      1 & e^{i\theta}  & -1 & -e^{i\theta}\\
      1 & -1 & 1 & -1\\
      1 & -e^{i\theta} & -1 & e^{i\theta} 
\end{array} \right),
\end{equation} 
which reduces to the canonical case, Eq.~(\ref{eq:matrixelements}), when $\theta=\pi/2$. 
The existence of these free internal phases for $N\geq 4$ provides an extra degree of freedom for making a wide range of nonclassical states using multiport beamsplitters with single photon and coherent state inputs. An open experimental question is how close the transfer matricies of fused fibre multiports come to the canonical matrix, and whether production can be tuned to obtain the canonical case. 

Finally, there exists significant freedom to explore other interesting nonclassical states that may be produced by techniques similar to the protocol we have presented. We have already mentioned varying the internal phases of the $2N$-port device, as well as the use of more than one coherent state input, but one can also modify the output network, the postselection protocol, or the input states \cite{olefock}. An instance of the latter is using $n>1$ Fock states. For example if the input state of an 8-port is set to $|\psi\rangle_{A}=|2,2,1,1\rangle$, and single photons are detected in the $b_{0}$ and $b_{2}$ modes, then in the modes $b_{1}$ and $b_{3}$ the state is {\em exactly} the NOON state, $|\psi\rangle_{B}=|4,0 \rangle + |0,4 \rangle$, without the need for postselection or zero photon detections. This success probability of this configuration is 3/64, the same as the more complicated circuit in Ref. \cite{kok}.

In conclusion, we have theoretically demonstrated a technique for producing maximally entangled photon number states using $2N$-port fiber splitters, photon-counting postselection or conditional measurement, and Fock-state inputs. We have also shown that it is possible to replace one of the inputs with a coherent state and achieve the same maximally entangled state. This technique allows a highly nonlinear state transformation to be realized experimentally with a simple optical circuit. 

{\em Note added.} Recently we became aware of an independent proposal for implementing phase measurements and nonlinear gates using multiport devices and postselection \cite{steuernagel}.

Many thanks to G. J. Milburn for lots of helpful and stimulating discussions, and for the proof below. We also wish to thank A. Gilchrist and T. C. Ralph for helpful discussions. This work was supported by the Australian Research Council, and by the National Security Agency (NSA) and the Advanced Research and Development Activity (ARDA) under Army Research Office (ARO) contract number DAAD-19-01-1-0651.

\appendix*
\section{}
We prove the result in Eq.~(\ref{eq:psstate}), with complex numbers $\beta$, $\gamma$ replacing $\hat{b}_0^\dag$,   $\hat{b}_1^\dag$, since the latter commute.
\begin{equation}
\prod_{k=0}^{N-1}(\beta+e^{2\pi i k/N}\gamma)=\beta^N-(-1)^N\gamma^N
\label{product}
\end{equation}
%\vskip 1 truecm

\noindent {\em Proof.} Define the set of numbers $\lambda_k=\beta+e^{2\pi i k/N}\gamma$. These numbers  are the roots of
\begin{equation}
(\beta-\lambda_k)^N-(-1)^N\gamma^N=0
\label{character}
\end{equation}
which is in fact the characteristic equation for the matrix
\begin{equation}
M=\left (\begin{array}{ccccc}
  \beta & \gamma & 0 & \ldots  & 0\\
 0 & \beta & \gamma & \ldots & 0\\
\vdots &\vdots & & & \\
\gamma & 0 & \ldots & 0 & \beta
\end{array}\right )
\end{equation}
Thus we see that the left hand side of Eq.~(\ref{product}) is in fact the product of the eigenvalues for $M$, in other words it is the determinant of $M$. The determinant of $M$ is simply the left hand side of the characteristic equation, Eq.~(\ref{character}) , with $\lambda=0$. Thus Eq.~(\ref{product}) follows. 

%\vspace{-0.4 cm}


\begin{thebibliography}{99}
\expandafter\ifx\csname natexlab\endcsname\relax\def\natexlab#1{#1}\fi
\expandafter\ifx\csname bibnamefont\endcsname\relax
  \def\bibnamefont#1{#1}\fi
\expandafter\ifx\csname bibfnamefont\endcsname\relax
  \def\bibfnamefont#1{#1}\fi
\expandafter\ifx\csname citenamefont\endcsname\relax
  \def\citenamefont#1{#1}\fi
\expandafter\ifx\csname url\endcsname\relax
  \def\url#1{\texttt{#1}}\fi
\expandafter\ifx\csname urlprefix\endcsname\relax\def\urlprefix{URL }\fi
\providecommand{\bibinfo}[2]{#2}
\providecommand{\eprint}[2][]{\url{#2}}

%\vspace{-0.9 cm}
\bibitem[{\citenamefont{Dowling and Milburn}(2003)}]{dowlingmilburn}
\bibinfo{author}{\bibfnamefont{J.~P.} \bibnamefont{Dowling}} \bibnamefont{and}
  \bibinfo{author}{\bibfnamefont{G.~J.} \bibnamefont{Milburn}},
  \bibinfo{journal}{Philos. Trans. R. Soc. London, Ser. A} \textbf{\bibinfo{volume}{361}}, \bibinfo{pages}{1655}
  (\bibinfo{year}{2003}).

\bibitem[{\citenamefont{Knill et~al.}(2001)\citenamefont{Knill, Laflamme, and
  Milburn}}]{klm}
\bibinfo{author}{\bibfnamefont{E.}~\bibnamefont{Knill}},
  \bibinfo{author}{\bibfnamefont{R.}~\bibnamefont{Laflamme}}, \bibnamefont{and}
  \bibinfo{author}{\bibfnamefont{G.~J.} \bibnamefont{Milburn}},
  \bibinfo{journal}{Nature (London)} \textbf{\bibinfo{volume}{409}}, \bibinfo{pages}{46}
  (\bibinfo{year}{2001}), for example.

\bibitem[{\citenamefont{Gisin et~al.}(2002)\citenamefont{Gisin, Ribordy,
  Tittel, and Zbinden}}]{gisin}
\bibinfo{author}{\bibfnamefont{N.}~\bibnamefont{Gisin}}
   \bibnamefont{\emph{et al.}},
  \bibinfo{journal}{Rev. Mod. Phys.} \textbf{\bibinfo{volume}{74}},
  \bibinfo{pages}{145} (\bibinfo{year}{2002}).

\bibitem[{\citenamefont{Lee et~al.}(2002)\citenamefont{Lee, Kok, and
  Dowling}}]{dowlingrosetta}
\bibinfo{author}{\bibfnamefont{H.}~\bibnamefont{Lee}},
  \bibinfo{author}{\bibfnamefont{P.}~\bibnamefont{Kok}}, \bibnamefont{and}
  \bibinfo{author}{\bibfnamefont{J.~P.} \bibnamefont{Dowling}},
  \bibinfo{journal}{J. Mod. Opt.} \textbf{\bibinfo{volume}{49}},
  \bibinfo{pages}{2325} (\bibinfo{year}{2002}).

\bibitem[{\citenamefont{Edamatsu et~al.}(2002)\citenamefont{Edamatsu, Shimizu,
  and Itoh}}]{itoh}
\bibinfo{author}{\bibfnamefont{K.}~\bibnamefont{Edamatsu}},
  \bibinfo{author}{\bibfnamefont{R.}~\bibnamefont{Shimizu}}, \bibnamefont{and}
  \bibinfo{author}{\bibfnamefont{T.}~\bibnamefont{Itoh}},
  \bibinfo{journal}{Phys. Rev. Lett.} \textbf{\bibinfo{volume}{89}},
  \bibinfo{pages}{213601} (\bibinfo{year}{2002}).
  
  \bibitem[{\citenamefont{Fonseca et~al.}(2002)\citenamefont{Fonseca et al}}]{fonseca}
\bibinfo{author}{\bibfnamefont{E.~J.~S.}~\bibnamefont{Fonseca}} \bibnamefont{\emph{et al.}},
  \bibinfo{journal}{Phys. Rev. Lett.} \textbf{\bibinfo{volume}{82}},
  \bibinfo{pages}{2868} (\bibinfo{year}{1999}).
  
\bibitem[{\citenamefont{Bollinger et~al.}(1996)\citenamefont{Bollinger, Itano,
  Wineland, and Heinzen}}]{bollinger}
\bibinfo{author}{\bibfnamefont{J.~J.} \bibnamefont{Bollinger}} \bibnamefont{\emph{et al.}}, \bibinfo{journal}{Phys. Rev. A}
  \textbf{\bibinfo{volume}{54}} (\bibinfo{year}{1996}).

\bibitem[{\citenamefont{Ralph}(2002)}]{ralph}
\bibinfo{author}{\bibfnamefont{T.~C.} \bibnamefont{Ralph}},
  \bibinfo{journal}{Phys. Rev. A} \textbf{\bibinfo{volume}{65}},
  \bibinfo{pages}{042313} (\bibinfo{year}{2002}).

\bibitem[{\citenamefont{Boto et~al.}(2000)\citenamefont{Boto, Kok, Abrams,
  Braunstein, Williams, and Dowling}}]{boto}
\bibinfo{author}{\bibfnamefont{A.~N.} \bibnamefont{Boto}}
   \bibnamefont{\emph{et al.}}, \bibinfo{journal}{Phys. Rev. Lett.}
  \textbf{\bibinfo{volume}{85}}, \bibinfo{pages}{2733} (\bibinfo{year}{2000}).

\bibitem[{\citenamefont{D\mbox{'}Angelo
  et~al.}(2001)\citenamefont{D\mbox{'}Angelo, Chekhova, and Shih}}]{shihlithog}
\bibinfo{author}{\bibfnamefont{M.}~\bibnamefont{D\mbox{'}Angelo}},
  \bibinfo{author}{\bibfnamefont{M.~V.} \bibnamefont{Chekhova}},
  \bibnamefont{and} \bibinfo{author}{\bibfnamefont{Y.}~\bibnamefont{Shih}},
  \bibinfo{journal}{Phys. Rev. Lett.} \textbf{\bibinfo{volume}{87}},
  \bibinfo{pages}{013602} (\bibinfo{year}{2001}).

\bibitem[{\citenamefont{Strekalov et~al.}(1995)\citenamefont{Strekalov,
  Sergienko, Klyshko, and Shih}}]{imag}
\bibinfo{author}{\bibfnamefont{D.~V.} \bibnamefont{Strekalov}}
   \bibnamefont{\emph{et al.}},
  \bibinfo{journal}{Phys. Rev. Lett.} \textbf{\bibinfo{volume}{74}},
  \bibinfo{pages}{3600} (\bibinfo{year}{1995}).

\bibitem[{\citenamefont{Meyer et~al.}(2001)\citenamefont{Meyer, Rowe,
  Kielpinski, Sackett, Itano, Monroe, and Wineland}}]{meyer}
\bibinfo{author}{\bibfnamefont{V.}~\bibnamefont{Meyer}}
  \bibnamefont{\emph{et al.}},
  \bibinfo{journal}{Phys. Rev. Lett.} \textbf{\bibinfo{volume}{86}},
  \bibinfo{pages}{5870} (\bibinfo{year}{2001}).

\bibitem[{\citenamefont{Fiur\'{a}\v{s}ek}(2002)}]{fiurasek}
\bibinfo{author}{\bibfnamefont{J.}~\bibnamefont{Fiur\'{a}\v{s}ek}},
  \bibinfo{journal}{Phys. Rev. A} \textbf{\bibinfo{volume}{65}},
  \bibinfo{pages}{053818} (\bibinfo{year}{2002}).

\bibitem[{\citenamefont{Kok et~al.}(2002)\citenamefont{Kok, Lee, and
  Dowling}}]{kok}
\bibinfo{author}{\bibfnamefont{P.}~\bibnamefont{Kok}} \bibnamefont{\emph{et al.}},
  \bibinfo{journal}{Phys. Rev. A} \textbf{\bibinfo{volume}{65}},
  \bibinfo{pages}{052104} (\bibinfo{year}{2002}).

\bibitem[{\citenamefont{Zou et~al.}(2001)\citenamefont{Zou, Pahlke, and
  Mathis}}]{zou}
\bibinfo{author}{\bibfnamefont{X.}~\bibnamefont{Zou}}
   \bibnamefont{\emph{et al.}},
  \bibinfo{journal}{quant-ph/0110149}  (\bibinfo{year}{2001}).

\bibitem[{\citenamefont{Gerry and Campos}(2001)}]{gerry}
\bibinfo{author}{\bibfnamefont{C.~C.} \bibnamefont{Gerry}} \bibnamefont{and}
  \bibinfo{author}{\bibfnamefont{R.~A.} \bibnamefont{Campos}},
  \bibinfo{journal}{Phys. Rev. A} \textbf{\bibinfo{volume}{64}},
  \bibinfo{pages}{063814} (\bibinfo{year}{2001}).
  
  \bibitem{note2N}\bibnamefont{Although this device consists of $N$ modes, we use the term ``$2N$-port'' \cite[for example]{mattle, zukowski} to identify the fact that there are two ports (one input and one output) associated with each mode. An optical fibre $2N$-port will have $2N$ fibre connectors.}

\bibitem[{\citenamefont{Mattle et~al.}(1995)\citenamefont{Mattle, Michler,
  Weinfurter, Zeilinger, and Zukowski}}]{mattle}
\bibinfo{author}{\bibfnamefont{K.}~\bibnamefont{Mattle}}
  \bibnamefont{\emph{et al.}},
  \bibinfo{journal}{Appl. Phys. B} \textbf{\bibinfo{volume}{60}}
 \bibinfo{pages}{S111} (\bibinfo{year}{1995}).
  
  \bibitem{zukowski}
  \bibinfo{author}{\bibnamefont{M.}~\bibnamefont{Zukowski}}
   \bibnamefont{\emph{et al.}},
  \bibinfo{journal}{Phys. Rev. A} \textbf{\bibinfo{volume}{55}},
 \bibinfo{pages}{2464} (\bibinfo{year}{1997}).
 

\bibitem[{\citenamefont{Takeuchi et~al.}(1999)\citenamefont{Takeuchi, Kim,
  Yamamoto, and Hogue}}]{takeuchi}
\bibinfo{author}{\bibfnamefont{S.}~\bibnamefont{Takeuchi}}
  \bibnamefont{\emph{et al.}},
  \bibinfo{journal}{Appl. Phys. Lett.} \textbf{\bibinfo{volume}{74}},
  \bibinfo{pages}{1063} (\bibinfo{year}{1999}).

\bibitem[{\citenamefont{Kok and Braunstein}(2001)}]{kokbraunstein}
\bibinfo{author}{\bibfnamefont{P.}~\bibnamefont{Kok}} \bibnamefont{and}
  \bibinfo{author}{\bibfnamefont{S.~L.} \bibnamefont{Braunstein}},
  \bibinfo{journal}{Phys. Rev. A} \textbf{\bibinfo{volume}{63}},
  \bibinfo{pages}{033812} (\bibinfo{year}{2001}).

\bibitem[{\citenamefont{Jennewein et~al.}(2002)\citenamefont{Jennewein, Weihs,
  Pan, and Zeilinger}}]{jennewein}
\bibinfo{author}{\bibfnamefont{T.}~\bibnamefont{Jennewein}}
  \bibnamefont{\emph{et al.}},
  \bibinfo{journal}{Phys. Rev. Lett.} \textbf{\bibinfo{volume}{88}},
  \bibinfo{pages}{017903} (\bibinfo{year}{2002}).

\bibitem[{\citenamefont{Bernstein}(1974)}]{bernstein}
\bibinfo{author}{\bibfnamefont{H.~J.} \bibnamefont{Bernstein}},
  \bibinfo{journal}{J.Math. Phys.} \textbf{\bibinfo{volume}{15}},
  \bibinfo{pages}{1677} (\bibinfo{year}{1974}).

\bibitem[{\citenamefont{Steuernagel}(1997)}]{olefock}
\bibinfo{author}{\bibfnamefont{O.}~\bibnamefont{Steuernagel}},
  \bibinfo{journal}{Opt. Commun.} \textbf{\bibinfo{volume}{138}},
  \bibinfo{pages}{71} (\bibinfo{year}{2003}).

\bibitem[{\citenamefont{Steuernagel}(2003)}]{steuernagel}
\bibinfo{author}{\bibfnamefont{O.}~\bibnamefont{Steuernagel}},
  \bibinfo{journal}{quant-ph/0303068}  (\bibinfo{year}{2003}).

\end{thebibliography}
\end{document}